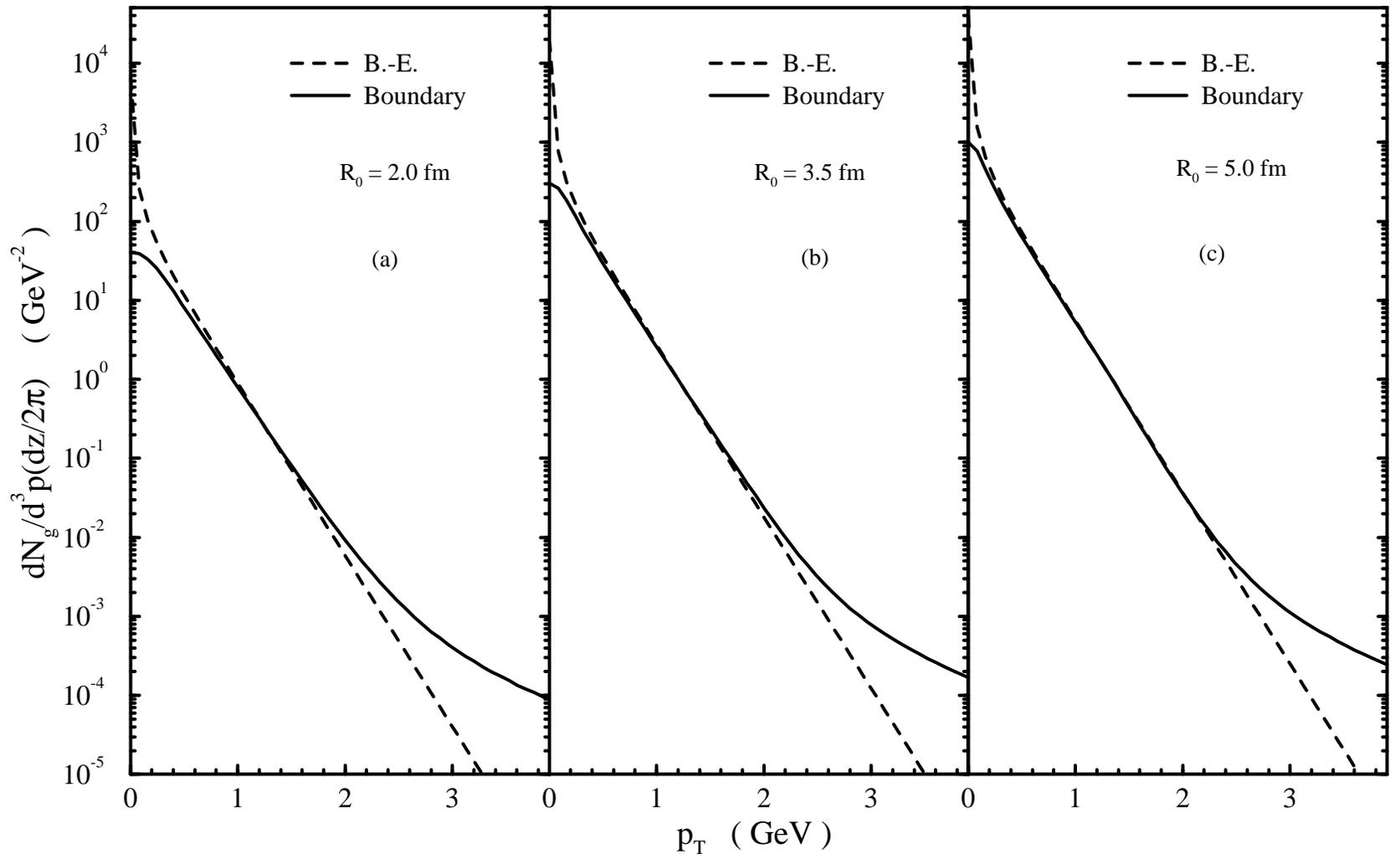

Fig.(1)

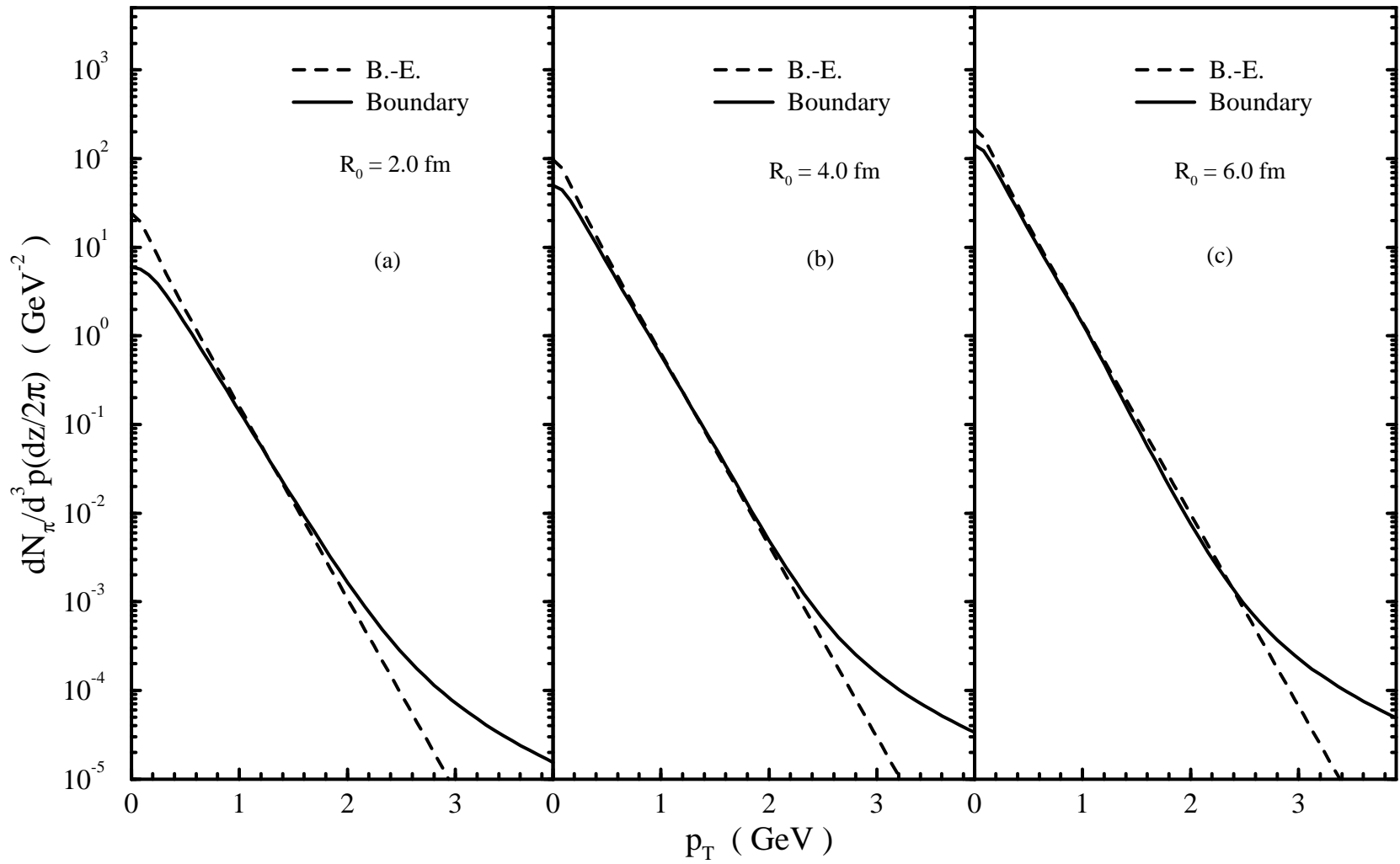

Fig.(2)

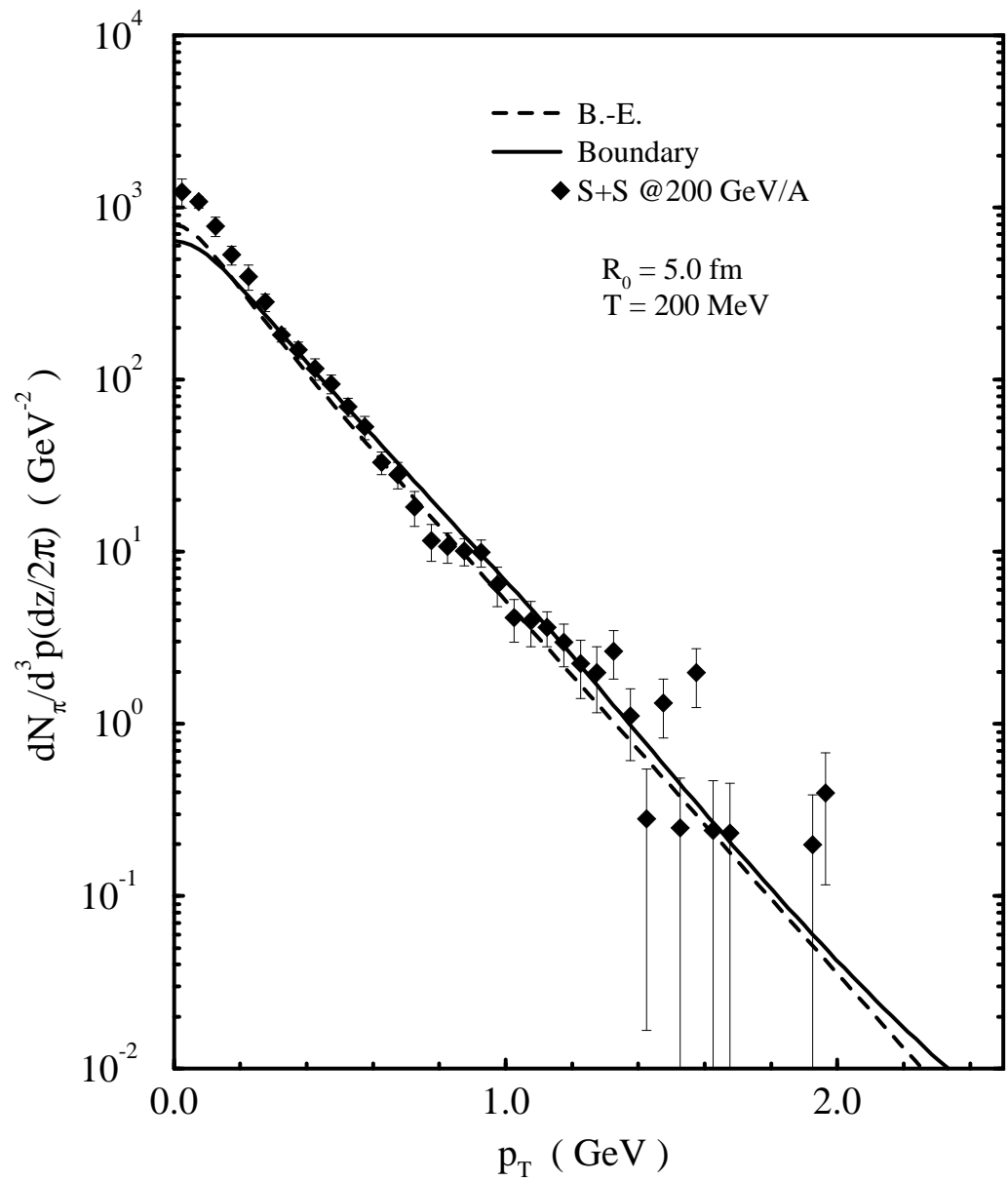

fig.(3)

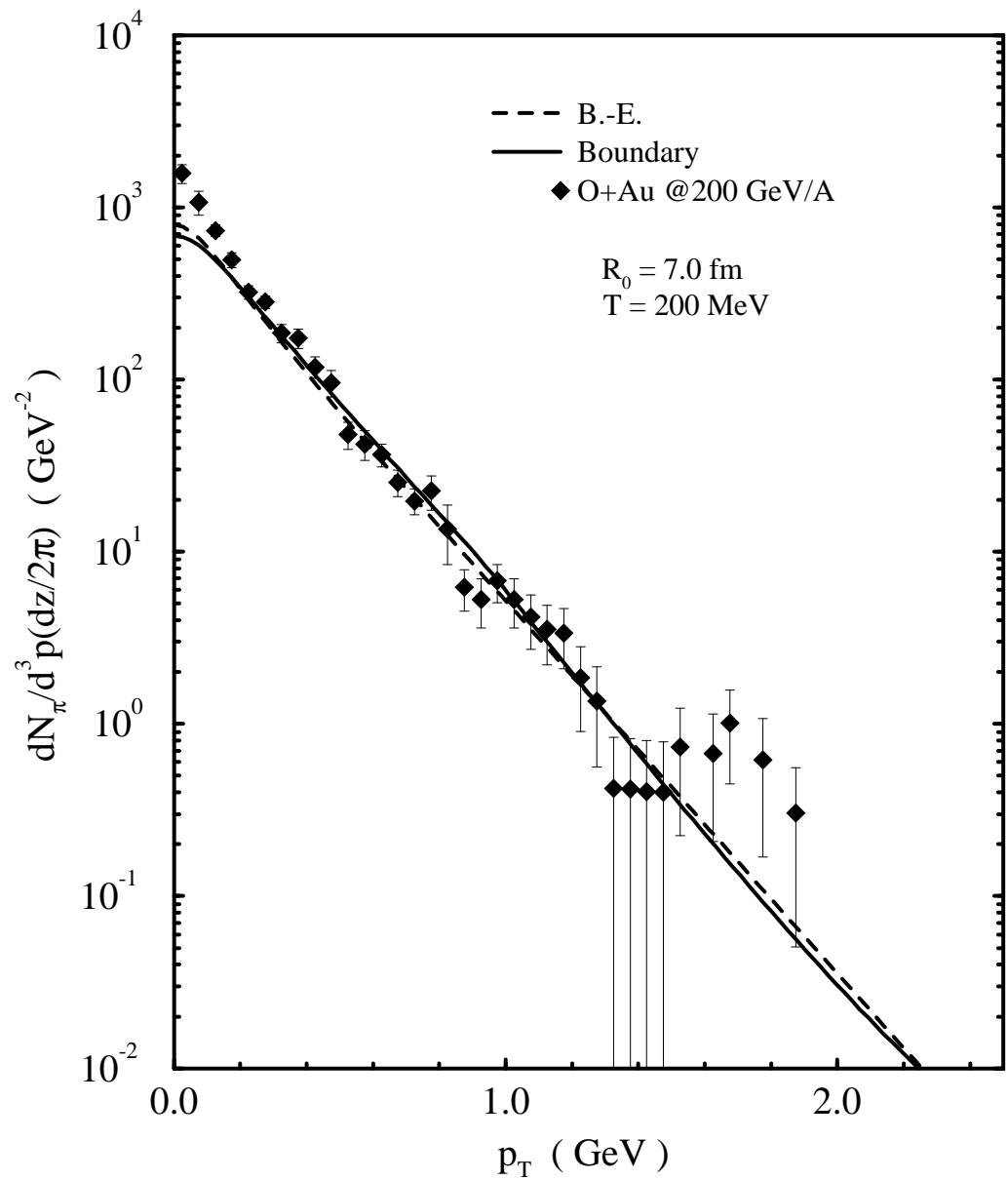

Fig.(4)

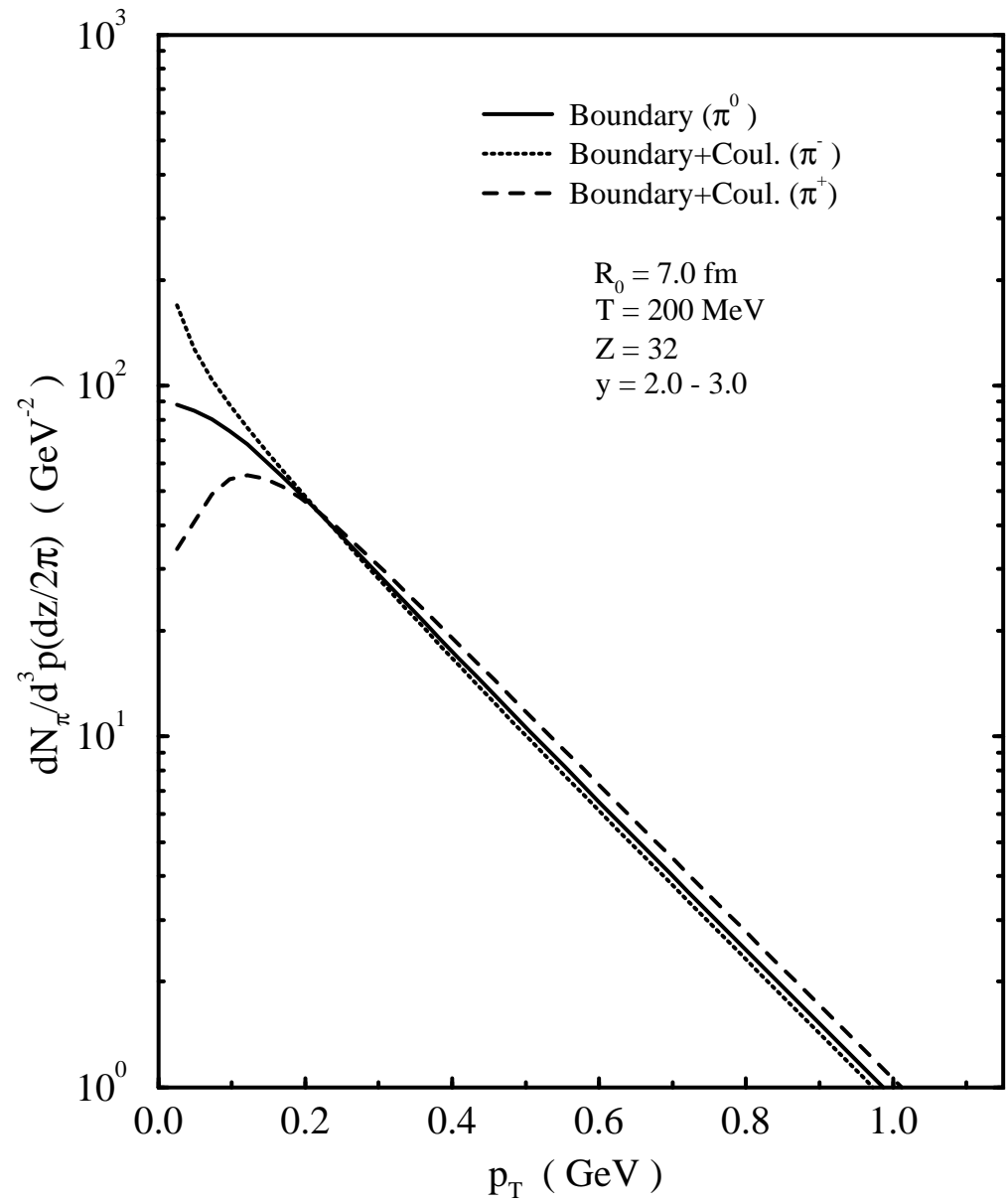

Fig.(5)

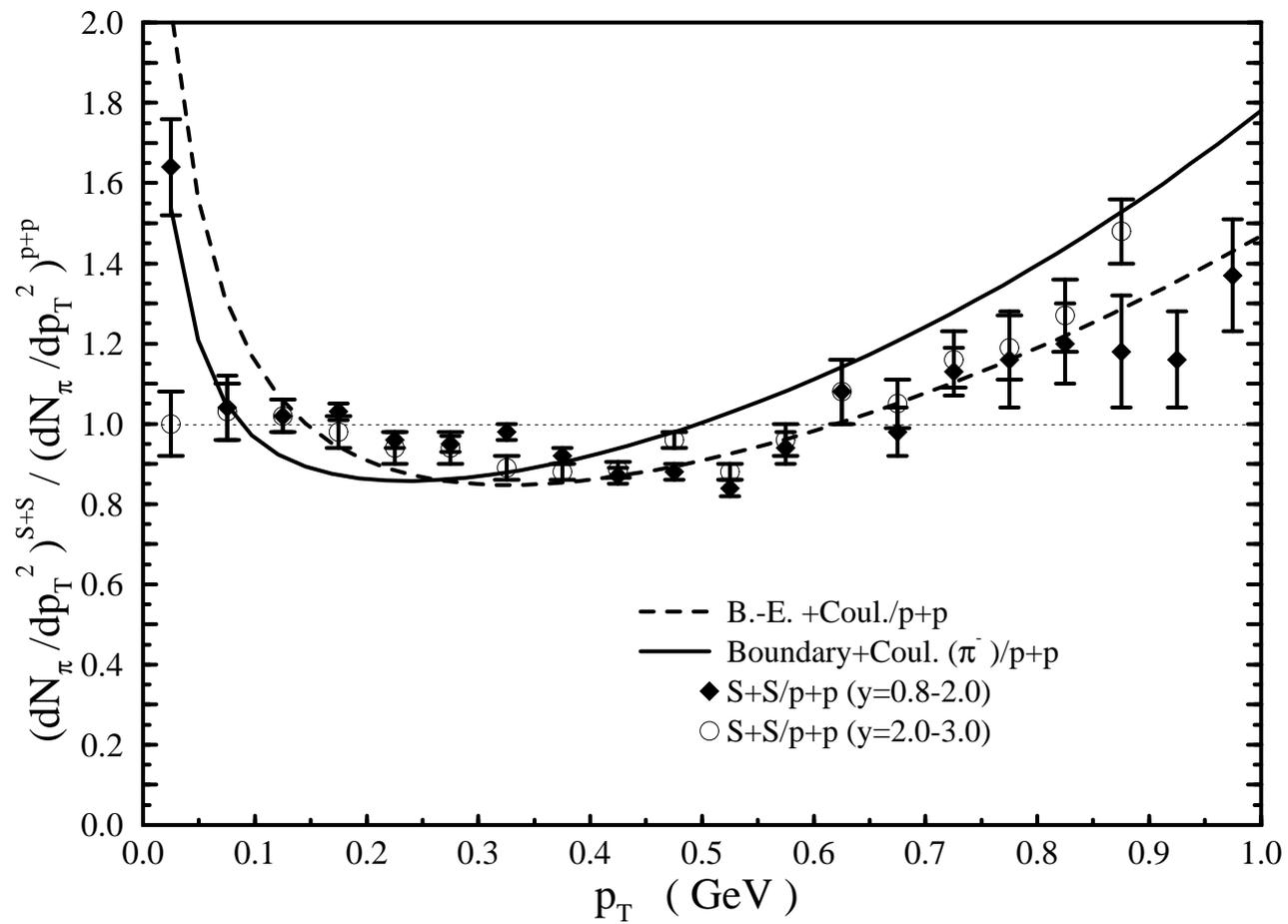

Fig.(6)

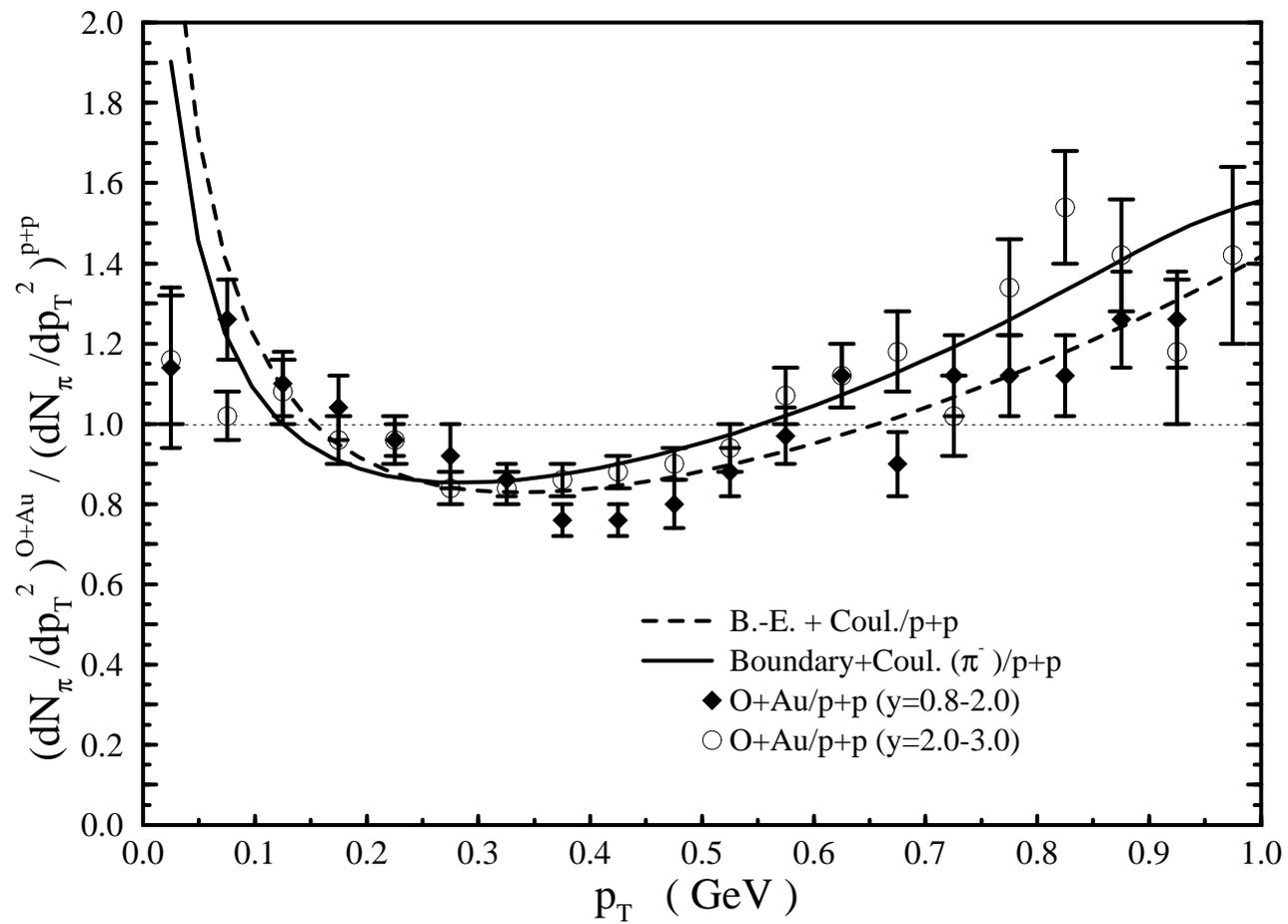

Fig.(7)

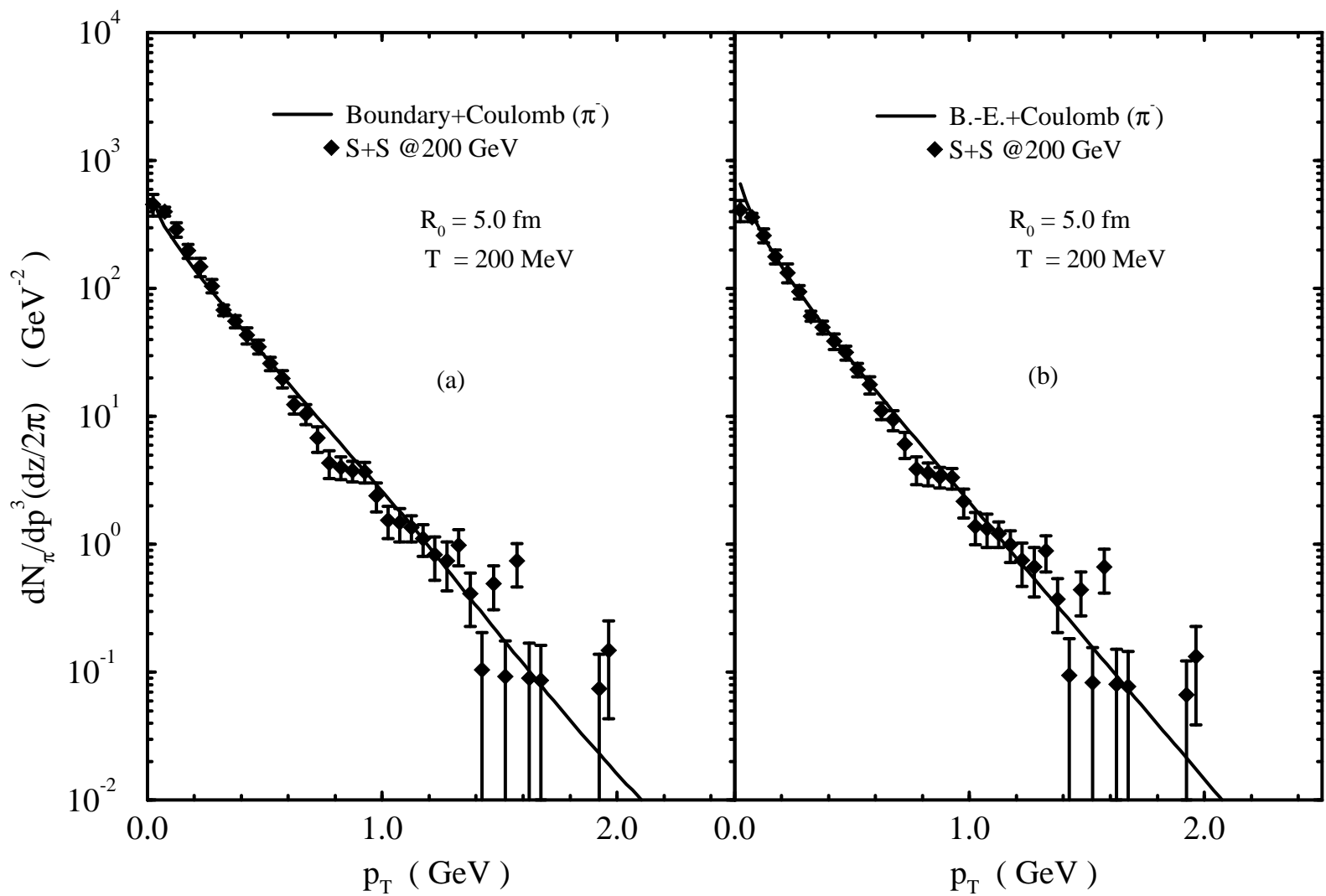

Fig.8

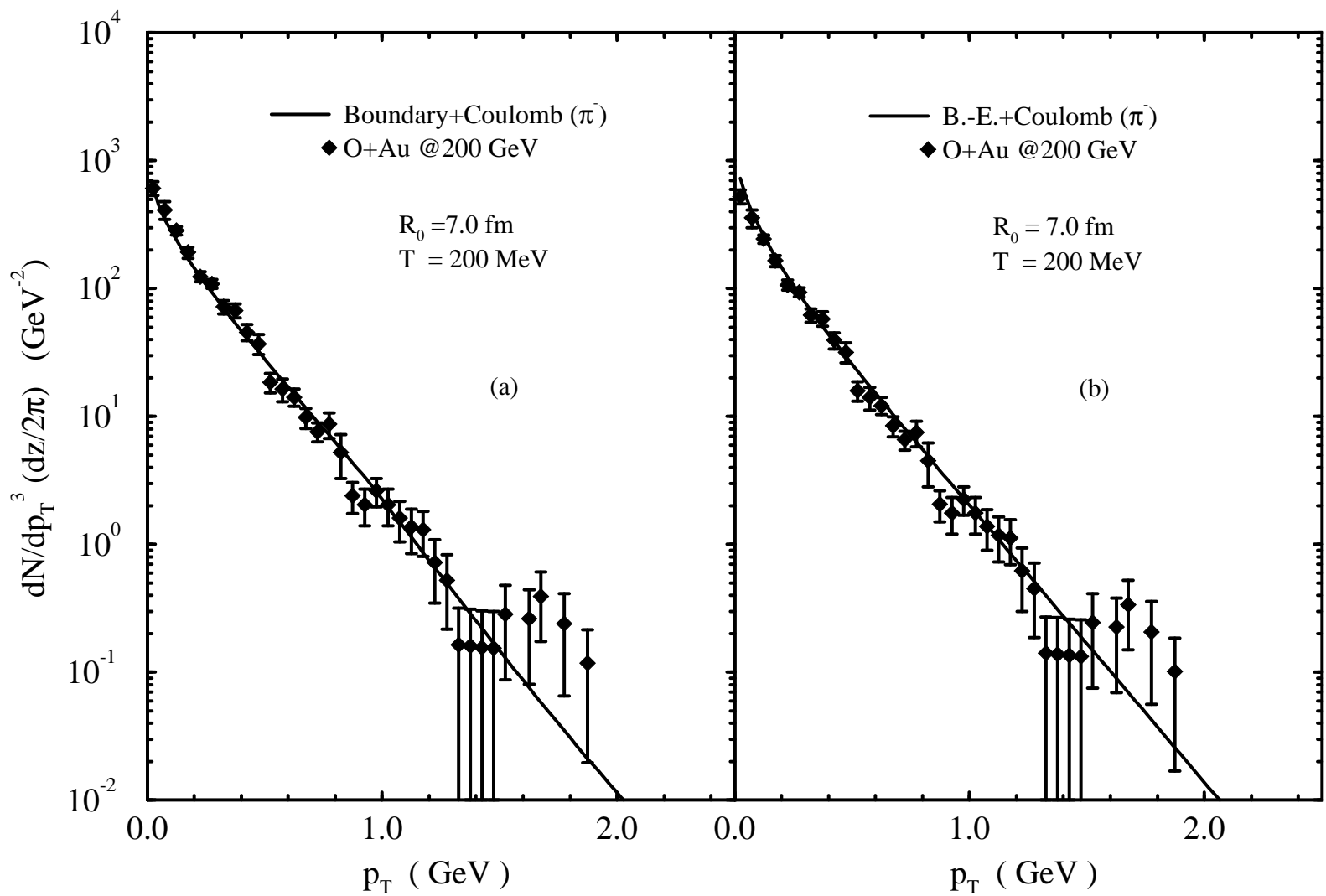

Fig.9

# Boundary and Coulomb Effects on Boson Systems in High-Energy Heavy-Ion Collisions

M. G.-H. Mostafa[1,2][*] and Cheuk-Yin Wong[1][†]

[1] *Oak Ridge National Laboratory, Oak Ridge, TN 37831-6373, USA*

and

[2] *Physics Department, Faculty of Science, Ain Shams University, Cairo, Egypt.*

(December 22, 1994)

## Abstract

The boundary of a boson system plays an important role in determining the momentum distribution of the bosons. For a boson system with a cylindrical boundary, the momentum distribution is enhanced at high transverse momenta but suppressed at low transverse momenta, relative to a Bose-Einstein distribution. The boundary effects on systems of massless gluons and massive pions are studied. For gluons in a quark-gluon plasma, the presence of the boundary may modify the signals for the quark-gluon plasma. For pions in a pion system in heavy-ion collisions, Coulomb final-state interactions with the nuclear participants in the vicinity of the central rapidity region further modify the momentum distribution at low transverse momenta. By including both the boundary effect and the Coulomb final-state interactions we are able to account for the behavior of the $\pi^-$ transverse momentum spectrum observed in many heavy-ion experiments, notably at low transverse momenta.

Typeset using REVTEX 3.0

---

[*]E-Mail: mostafa@orph01.phy.ornl.gov (gadalhaq@frcu.eun.eg, after June, 1995).

[†]E-Mail: wong@orph01.phy.ornl.gov.

# I. INTRODUCTION

Recently, the production of high-density and high-temperature matter in relativistic nuclear collisions has received much attention. It is generally expected that these collisions may provide the tools to probe the existence of a new phase of matter of strongly interacting particles, the quark-gluon plasma (QGP), at high temperature or at high baryon density [1]. In the QGP, quarks and gluons are deconfined; the deconfinement refers to a circumstance in which quarks, anti-quarks, and gluons are no longer confined within the spatial dimensions of a hadron. Quarks and gluons are nonetheless confined within the boundary of the plasma. Under laboratory conditions in which a QGP is produced using relativistic heavy-ion collisions, the expected initial transverse radius of the QGP is comparable to the smaller of the radii of the two colliding nuclei producing the plasma, typically a few fermis. Therefore, a corollary of the existence of a QGP in a heavy-ion collision is that the quarks and gluons in the plasma travel freely within a spatial region of a few fermis.

The distribution of transverse momentum provides information on the transverse boundary of the plasma. It can be used to detect the presence of the QGP. Furthermore, the magnitudes of the signals for QGP detection are functions of the momentum distribution of the quarks and gluons, which has generally been taken to be a thermal distribution [2–4]. Because the presence of the boundary affects the momentum distribution [5], the boundary effects should be taken into account to give better estimates of the magnitude of the signals of the quark-gluon plasma. While the boundary effects on quarks have been studied previously [5], here we wish to study the effect of the boundary on the distribution of the transverse momenta of gluons in a QGP, using wave functions which satisfy the boundary conditions.

Another interesting boson system in which boundary effects may manifest themselves is the multipion system produced after a high energy heavy-ion collision. If the mean-field potential of the pion system has a boundary before it freezes-out, the transverse momentum distribution will show the effect of the boundary.

The subject of the transverse momentum distribution of pions has received much attention recently. The latest experimental data [6–8] show that the transverse momentum distributions of negative hadrons, mainly pions, from hadron-nucleus and nucleus-nucleus collisions exhibit a striking enhancement at low and high transverse momenta relative to that of negative hadrons from $p+p$ collisions at comparable energies. This origin of this puzzling phenomenon has been vigorously debated in many articles [9–17]. Atwater *et al.* [9] and Lee and Heinz [10] interpreted the data by assuming a transversely expanding fireball. With assumed velocity profiles, Kusnezov and Bertsch [11] argued that the transverse velocity profiles used in references [9,10] imply an unnatural assumption concerning the nature of the freeze-out surface. Furthermore, they indicated that a more complete hydrodynamical description of the transverse flow cannot account for the experimental shape of the pion transverse momentum distribution at low $p_T$. Another explanation, suggested by several authors [12–14], is a non-zero chemical potential for the pions. Alternatively, Sollfrank, Koch, and Heinz [15] found agreement with the S+S data by including contributions due to resonance decays on hadronic momentum spectra. However, Barz *et al* [16] examined the role of decays of the excited hadrons in determining the pion $p_T$ spectra, and argued that no reasonable statistical model can reproduce the experimentally observed peak at low $p_T$ in the pion transverse momentum distributions. A very good agreement with the S+S data has also been found by Bolz et al [17] by modifying Landau initial conditions with an initial rapidity distribution which is only a function of the longitudinal coordinate.

In this work we study the boundary and Coulomb effects on pion systems at high temperature. The momentum distribution of particles in a bounded medium is altered because the single-particle wave function must satisfy the appropriate boundary conditions. The pions produced after a nucleus-nucleus collision are moving in the Coulomb field of the colliding nuclear participants, whose charge depends on the impact parameter. For a head-on collision, the charge of the colliding participants in the central rapidity region can be as big as $2Z$ for two identical nuclei. The momentum distribution of charged pions is changed due to the Coulomb final-state interactions between the pions and the protons near the center of mass region. This Coulomb effect depends on the factor $\eta = Z\alpha/v$, which is large in heavy-ion central collisions. This final state interaction has a large effect on the charged pions with small transverse momenta, as shown in this paper.

This paper is organized as follows: In Section II, we describe the method for obtaining the momentum distribution of a boson system. In Section III, the transverse momentum distribution of gluons in a quark-gluon plasma is obtained, using wave functions satisfying the boundary conditions. In Section IV, we study the boundary effect on the pion transverse momentum distribution in a pion system at high temperatures. The effect of the Coulomb final-state interaction on the pion transverse momentum distribution is discussed in Section V. Finally, Section VI contains the summary and conclusions.

## II. MOMENTUM DISTRIBUTION OF A BOSON SYSTEM

For a boson system in a mean-field scalar potential $m(\vec{x})$, if $\tilde{\Phi}_\lambda(\mathbf{p})$ is the boson single-particle wave function in a quantum state $\lambda$ in momentum space, the momentum distribution of bosons is given by:

$$\frac{dN_B}{d^3\mathbf{p}} = g_B \sum_\lambda W(E_\lambda) \, |\tilde{\Phi}_\lambda(\mathbf{p})|^2 , \qquad (1)$$

where the summation is carried out over all the single-particle quantum numbers $\{\lambda\}$, $g_B$ is the boson degeneracy number, and $W(E_\lambda)$ is the occupation probability of the single-particle state $\lambda$. The thermodynamical state of the system is described by the distribution $W(E_\lambda)$, which is a function of the single-particle energy $E_\lambda$. For a boson system at thermal equilibrium $W(E)$ is represented by the Bose-Einstein distribution, which for a boson system at temperature $T$ and chemical potential $\mu$ is given by

$$W(E) = [e^{(E-\mu)/T} - 1]^{-1} .$$

The wave function $\tilde{\Phi}(\mathbf{p})$, for a boson system with a cylindrical boundary, is given by the Fourier transform

$$\tilde{\Phi}(\mathbf{p}) = \frac{1}{(2\pi)^{3/2}} \int e^{i\mathbf{r}\cdot\mathbf{p}} \, \Phi(r, \phi, z) \, d\vec{r} ,$$

where $\Phi(r, \phi, z)$ is the corresponding single-particle wave function in configuration space, which is obtained by solving the Klein-Gordon equation for a boson in a mean-field potential $m(r)$. The Klein-Gordon equation, for the wave function $\Phi(r,\phi,z)$ of a boson in a boson system with a cylindrical boundary and moving in a mean-field scalar potential $m(r)$ varying in the transverse direction, is given by

$$\left[p^2 - m^2(r)\right]\Phi(r,\phi,z) = 0, \tag{2}$$

where $p$ is the boson four-momentum. Different boson systems will be described by different mean-field potentials $m(r)$.

## III. BOUNDARY EFFECTS ON A GLUON PLASMA

We consider a gluon in a quark-gluon plasma in the form of a sharp cylinder, which is presumed to be formed in a high energy nucleus-nucleus collision. The mean-field potential $m(r)$ is defined by:

$$m(r) = \begin{cases} m_g & \text{if } r \leq R_o \text{ (region I)} \\ M & \text{if } r > R_o \text{ (region II)}, \end{cases}$$

for gluons we set $m_g = 0$. Separation of the wave function $\Phi(r,\phi,z)$ in Eq. (2) into

$$\Phi(r,\phi,z) = R_\nu(r)e^{\pm i\nu\phi}Z(z)$$

leads to the longitudinal wave equation

$$\left[-\frac{d^2}{dz^2} + m^{*2} - E^2\right]Z(z) = 0 \tag{3}$$

and the radial wave equation

$$\left[\frac{d^2}{dr^2} + \frac{1}{r}\frac{d}{dr} + \left(C_\nu^2 - \frac{\nu^2}{r^2}\right)\right]R_\nu(r) = 0, \tag{4}$$

where $C_\nu^2 = m^{*2} - m^2(r)$, and $m^*$ is separation constant which physically is the transverse mass. For simplicity we shall study the case of an infinitely long cylinder; this gives $E^2 = P_{n_z}^2 + m^{*2}$, where $E$ is the eigenenergy and $\{P_{n_z}^2\}$ are the eigenvalues associated with the longitudinal wave functions $Z_{n_z}(z) = Ae^{ip_{n_z}z}$. Eq. (4) is a Bessel equation which has the following two solutions:

$$R_\nu^I(r) = AJ_\nu(C_\nu r) \quad \text{in region I}, $$
$$R_\nu^{II}(r) = A'K_\nu(C'_\nu r) \quad \text{in region II},$$

where $C_\nu^2 = m^{*2} - m_g^2$, $C_\nu'^2 = M^2 - m^{*2}$, and $A$ and $A'$ are overall normalizations. Using the continuity of the logarithmic derivatives of the two wave functions $R_\nu^I(r)$ and $R_\nu^{II}(r)$, and taking the asymptotic forms of both $J_\nu$ and $K_\nu$ as $M \to \infty$ in region II, we obtain an eigenvalue equation for $C_\nu$ as

$$R_\nu(R_o) = J_\nu(C_{\nu,s}R_o) = 0, \tag{5}$$

which gives $C_{\nu,s} = j_{\nu,s}/R_o$, where the values $\{j_{\nu,s}\}$ are the zeros of the Bessel function $J_\nu$. The radial wave function is then

$$R_{\nu,s}(r) = A J_\nu(C_{\nu,s} r).$$

We note that for the case of transverse confinement, $m^*$ is quantized and contributes a zero-point energy associated with motion in the transverse direction. Thus, as is evident from Eq. (3), even for a gluon with zero rest-mass, the gluon acquires a non-zero intrinsic momentum distribution due to the kinetic energy of confinement.

Introducing the Fourier transforms of $Z_{n_z}(z)$ and $R_{\nu,s}(r)e^{\pm i\nu\phi}$, we find longitudinal and transverse wave functions in momentum space,

$$\mathcal{F}_{n_z}(p_z) = \frac{1}{\sqrt{2\pi}} \int e^{ip_z z} Z_{n_z}(z) dz \qquad (6)$$

and

$$\begin{aligned}\tilde{\phi}_{\nu,s}(\mathbf{p}_T) &= \frac{1}{2\pi} \int d^2\mathbf{r}\, e^{i\mathbf{p}_T \cdot \mathbf{r}} R_{\nu,s}(r)\, e^{\pm i\nu\phi} \\ &= B \frac{1}{2\pi} \int r dr\, J_\nu(p_T r) J_\nu(C_{\nu,s} r), \end{aligned} \qquad (7)$$

where $B$ is a constant. Carrying out the integration gives [18]

$$\tilde{\phi}_{\nu,s}(\mathbf{p}_T) = \mathcal{N}_\nu^{-1/2} \frac{R_o}{p_T^2 - C_{\nu,s}^2} [p_T\, J_{\nu+1}(p_T R_o)\, J_\nu(C_{\nu,s} R_o) \\ - C_{\nu,s}\, J_\nu(p_T R_o)\, J_{\nu+1}(C_{\nu,s} R_o)], \qquad (8)$$

where

$$\mathcal{N}_\nu = \int d^2\mathbf{p}_T \left| \frac{R_o}{p_T^2 - C_{\nu,s}^2} [p_T\, J_{\nu+1}(p_T R_o)\, J_\nu(C_{\nu,s} R_o) \\ - C_{\nu,s}\, J_\nu(p_T R_o)\, J_{\nu+1}(C_{\nu,s} R_o)] \right|^2.$$

The wave function $\tilde{\phi}_{\nu,s}(\mathbf{p}_T)$ varies approximately as $p_T^{-3}$, with a small oscillatory component, at large $p_T$. The single-particle wave function in momentum space is defined as

$$\tilde{\Phi}_{n_z,\nu,s}(\mathbf{p}) = \tilde{\phi}_{\nu,s}(\mathbf{p}_T) \mathcal{F}_{n_z}(p_z),$$

and hence the gluon momentum distribution (Eq. (1)) becomes

$$\frac{dN_g}{d^3\mathbf{p}} = g_g \sum_{\nu s} \sum_{n_z} \frac{|\tilde{\phi}_{\nu,s}(\mathbf{p}_T)|^2 |\mathcal{F}_{n_z}(p_z)|^2}{e^{(E_{\nu,s} - \mu)/T} - 1}, \qquad (9)$$

where the gluon degeneracy number $g_g$ is 8. The transverse momentum distribution of the gluons in a quark-gluon plasma of temperature $T$ and chemical potential $\mu$ is therefore given by, using Eqs. (6), (8), and (9),

$$\frac{dN_g}{d^3\mathbf{p}(dz/2\pi)} = g_g \sum_{\nu,s} \left[e^{(E_{\nu,s}-\mu)/T} - 1\right]^{-1}$$
$$\times \mathcal{N}_\nu^{-1} \left|\frac{R_o}{p_T^2 - C_{\nu,s}^2}\left[p_T J_{\nu+1}(p_T R_o) J_\nu(C_{\nu,s} R_o) - C_{\nu,s} J_\nu(p_T R_o) J_{\nu+1}(C_{\nu,s} R_o)\right]\right|^2. \quad (10)$$

In contrast, if the boundary of the quark-gluon plasma extends to infinity, then the transverse momentum distribution is given by the Bose-Einstein form:

$$\frac{dN_g}{d^3\mathbf{p}(dz/2\pi)} = g_g \frac{\pi R_o^2}{(2\pi)^2} \frac{1}{e^{(E-\mu)/T} - 1}. \quad (11)$$

Fig. 1 shows the transverse momentum distributions of the gluons in a quark-gluon plasma of different radii $R_o = 2, 3.5,$ and $5$ fm, at temperature $T = 200$ MeV, $\mu = 0$, and $p_z = 0$. The solid curves give the gluon transverse momentum distributions including the boundary effects, as given by Eq. (10). The dashed curves give those distributions for an unconfined plasma, obtained using the Bose-Einstein distribution, given by Eq. (11). For a plasma of a small radius, the distribution given by Eq. (10) (the solid curves) differs from that given by the Bose-Einstein distribution of Eq. (11) (the dashed curves), as shown in Fig. 1(a). In each case, the distribution which includes the boundary effect is suppressed at low $p_T$ relative to the Bose-Einstein distribution; this is due to the small density of confined single-particle states at low energies. As the radius of the plasma increases, the distribution in the low momentum region increases and approaches the Bose-Einstein distribution of an unbounded boson system; this is shown in Fig. 1(b) and Fig. 1(c). The large difference between the two distributions at very low momenta, $p_T \leq 0.1$ GeV, is due to the fact that Bose-Einstein distribution diverges as the momentum of the massless gluon goes to zero, and in contrast the distribution with a boundary does not, due to the smaller low-$p_T$ density of states. In the opposite limit of large $p_T$, the bounded case has a large high-momentum tail, which varies with the transverse momentum by a power law, and exceeds the Bose-Einstein distribution by many orders of magnitude. This is due to the vanishing of the single-particle wave functions at the boundary of the plasma. The momentum at which the distribution begins to differ substantially from the Bose-Einstein distribution increases with the radius of the plasma.

In contrast to our results for the momentum distribution at low-$p_T$, Sinyukov [19] studied the finite-size effect on the momentum distribution of a massless particle using Bjorkan's scale hydrodynamics. He found that the single-particle transverse momentum at $p_T \approx 0$ is much enhanced over the Bose-Einstein distribution (The B.-E. distribution is infinite at $p_T \approx 0$ for an equilibrated massless particle gas). The physical reason of the enhancement is not known. His results also contradicts those of Bolz *et al.* [17] and Kataja *et al.* [20] who studied the transverse momentum distribution using Bjorkan's scale hydrodynamics. The $p_T$ distributions for massless particles in the latter two calculations do not show an enhancement over the B.-E. distribution.

## IV. EFFECT OF MEAN-FIELS BOUNDARY ON A PION SYSTEM

After an ultrarelativistic nucleus-nucleus collision the produced pions are at high temperature, typically 100 MeV to 200 MeV. It is well known that $\pi$-$\pi$ interaction in the isospin $I = 0$ state is attractive [21]. Furthermore, the strong interaction range of the pions is about 1.4 fm and the average separation between pions, $d$, in these pion systems can be estimated as follows. The pion density is approximately [22,5]

$$n = \frac{1}{\mathcal{A}} \frac{dN_\pi}{t_o dy}$$

where $t_o = 1$ fm/c is the pion production time and $\mathcal{A}$ is the transverse area of the collision region. For $^{16}$O on Au at $200A$ GeV, the peak value of $dN_\pi/dy$ is $120 \cdot \frac{3}{2} = 180$ and it is 120 for S on S at $200A$ GeV. The average separation, $d = n^{-1/3}$, between pions in the pion system produced in the former collision is about 0.6 fm and about 0.7 fm in the latter collision. Because the average separation between pions is smaller than the pion interaction range, the pions in such systems are strongly interacting with each other and they are moving in an attractive mean-field potential which extends over the pion system. The attractive mean-field leads to a quasi-bound pion system which can be considered to be in a liquid phase with a surface boundary, rather than in a gas phase. The notion of a pion liquid was introduced by Shuryak [23], to emphasize the importance of interparticle interactions; these interactions are attractive and have easily observable consequences.

We assume that hadronic matter after a heavy-ion collision is only confined within the boundary of their mean field when their average separation is smaller than the interaction range between pions. They become free pions when the pions' average separation in the system become larger than their interaction range. Because of the short range nature of the pion interactions, as the system expands, the transition from a pion with an attractive mean field potential to a free pion system, i.e. freeze-out, is a rapid process. The momentum distribution of the pions is governed much by their momentum distribution just before they freeze-out. It is therefore reasonable to use the boundary of their mean field before the freeze-out transition to calculate the momentum distribution of the observed pions. Pions in such a system have a momentum distribution that is modified by the presence of this boundary. For example, the boundary leads to a vanishing of the pion wave function throughout the exterior region, which enhances the momentum distribution in the high momenta region relative to a Bose-Einstein distribution. It is interesting to apply the previous formulation to this pion system, to develop our understanding of the boundary effects on a massive boson system.

We consider that a pion system in a cylinder with a sharp transverse (radial) boundary. We envision the boundary disappearing at the moment of freeze-out, while the momentum distribution of pions remain unchanged. The mean-field potential $m(r)$ for this pion system (before freeze-out) is taken to be

$$m(r) = \begin{cases} m_\pi & \text{if } r \leq R_o \text{ (region I)} \\ \infty & \text{if } r > R_o \text{ (region II)}, \end{cases}$$

where $m_\pi$ is the physical pion mass.

Following the discussion of section III, the pion transverse momentum distribution in the bounded system, with temperature $T$ and chemical potential $\mu$, is given by

$$\frac{dN_\pi}{d^3\mathbf{p}(dz/2\pi)} = g_\pi \sum_{\nu,s} \left[ e^{(E_{\nu,s} - \mu)/T} - 1 \right]^{-1}$$

$$\times \mathcal{N}_\nu^{-1} \left| \frac{R_o}{p_T^2 - C_{\nu,s}^2} [p_T\, J_{\nu+1}(p_T R_o)\, J_\nu(C_{\nu,s} R_o) - C_{\nu,s}\, J_\nu(p_T R_o)\, J_{\nu+1}(C_{\nu,s} R_o)] \right|^2 . \quad (12)$$

In contrast, the pion transverse momentum distribution without a boundary is given by the Bose-Einstein distribution,

$$\frac{dN_\pi}{d^3\mathbf{p}(dz/2\pi)} = g_\pi\, \frac{\pi R_o^2}{(2\pi)^2}\, \frac{1}{e^{(E-\mu)/T} - 1}, \quad (13)$$

where $E^2 = p_T^2 + p_z^2 + m_\pi^2$ and $g_\pi = 3$.

Fig. 2 shows the pion transverse momentum distributions arising with the boundary (using Eq. (12), solid curves) and with a Bose-Einstein Distribution (Eq. (13), dashed curves) for systems of different radii: $R_o = 2, 4,$ and 6 fm, $T = 200$ MeV, $p_z = 0$, and $\mu = 0$. The general features of the pion transverse momentum distribution, shown in Fig. 2, are similar to that of the gluons, shown in Fig. 1. For a pion system of small radius, the distribution arising from the boundary effect (solid curves) is quite different from that of a Bose-Einstein distribution (dashed curves) but they approach each other as the size of the system goes to infinity. Note however that the pion transverse momentum distribution at low momenta is broader than we found for gluons.

Fig. 3 shows the pion transverse momentum distribution Eq. (12) for a pion system with a boundary (solid curve), and the Bose-Einstein distribution Eq. (13) (dashed curve), compared with experimental data for S+S $\to negatives$ + X at $E = 200A$ GeV [6–8]. Fig. 4 similarly shows O+Au $\to negatives$ + X . The radius $R_o$ of the pion source is taken from the results of pion interferometry [24]. By equating the root-mean-square radius of our sharp distribution and the Gaussian distribution of [24], we have $R_o = \sqrt{2}\, R_T$, where $R_T$ is the effective transverse size of the source, as defined in [24]. Our calculations fit the experimental data well over most of $p_T$, given a pion system of radius $R_o = 5.0$ fm for S+S and $R_o = 7.0$ fm for O+Au, at $T = 200$ MeV, $\mu = 0$, and $p_z = 0$. However, at low momenta ($p_T \leq m_\pi$) neither the bounded pion system distribution Eq. (12) (solid curve) nor the Bose-Einstein distribution Eq. (13) (dashed curve) fits the data.

## V. COULOMB EFFECTS ON A PION SYSTEM

The Coulomb final-state interaction in relativistic nuclear collisions has been studied by many authors [25–27]. This effect has been found to play an important role in the momentum distribution of charged multiplicity produced in relativistic nucleus-nucleus collisions. For example, Libbrecht and Koonin [26] used a covariant classical formulation to show that a Coulomb focusing could arise toward $\theta_{c.m.} = 90°$ and finite $p_T$ in the Coulomb field generated by two charged fragments in relative motion (the projectile and the target remnants). Gyulassy and Kauffmann [27] used a relativistic field theoretic model to derive formulas to first order in $Z\alpha$ for the Coulomb final-state interactions. They derived covariant formulas that took into account multiple independently moving charged fragments of finite size and finite thermal expansion velocities, and showed the importance and complexity of the Coulomb final-state interaction in nuclear collisions.

The charged pion inclusive cross-section $\sigma_{\pi^\pm}(p)$, where $p$ is the particle momentum $|\vec{p}|$, to first order in $Z\alpha$ and to first order in the gradient of the pion-source current $J(x)$ is given (by Eq. (3.13) in [27]) as

$$\sigma_{\pi^\pm}(p) \approx \sigma_{\pi^\circ}(p \mp \delta k(p))\{1 \mp \delta D(p)\}, \tag{14}$$

where $\sigma_{\pi^\circ}(p)$ is the neutral pion inclusive cross-section, $\delta k(p)$ is the shift in the momentum $p$ due to the Coulomb final-state interaction, $\{1 \mp \delta D(p)\}$ is the Coulomb phase space factor, and the $\pm$ signs refers to the charge of the pion considered. In our calculations $\sigma_{\pi^\circ}(p)$ is identified either by the distribution of a bounded pion system, Eq. (12), or by a Bose-Einstein distribution, Eq. (13). The longitudinal momentum $p_z$, in terms of the rapidity $y$ and the transverse mass $m_T$, is

$$p_z = m_T \sinh y, \tag{15}$$

$$m_T = \sqrt{p_T^2 + m_\pi^2}. \tag{16}$$

The Coulomb phase space factor for a static charge density and in the limit of small momentum is found to be given by the Gamow factor $G(\eta)$, which is defined as:

$$G(\eta) = \frac{2\pi\eta}{e^{2\pi\eta} - 1}, \tag{17}$$

where for a $\pm|e|$ charged particle, $\eta = \pm Z\alpha/v$, where $Z$ is the total charge, $\alpha$ is the fine structure constant, and $v$ is the pion velocity. In our calculations we replace the first order expansion $\{1 \mp \delta D\}$ of Eq. (14) by the Gamow factor $G(\eta)$, to include Coulomb effect to all orders of $Z\alpha$. Eq. (14) thus becomes

$$\sigma_{\pi^\pm}(p) \approx \sigma_{\pi^\circ}(p \mp \delta k(p)) \frac{2\pi\eta}{e^{2\pi\eta} - 1}. \tag{18}$$

Fig. 5 shows how the Coulomb final-state interaction given by Eq. (18) affects the pion transverse momentum distribution, which was obtained previously using Eq. (12) for a pion system formed after a relativistic S+S collision, with $Z = 32$ and $R_o = 5.0$ fm. It is clear that, at low $p_T$, the Coulomb final-state interaction enhances the $\pi^-$ transverse momentum distribution and suppresses that of $\pi^+$ relative to $\pi^\circ$.

Fig. 6 shows, relative to that of a $p + p$ collision in comparable energies, the ratios of the normalized pion transverse momentum distributions for a bounded pion system given by Eq.(12) (solid curve), and the Bose-Einstein distribution of Eq.(13) (dashed curve), compared to experiment for S+S $\rightarrow$ *negatives* + X [28]. The Coulomb final-state interaction defined by Eq. (18) is taken into account, and calculations are carried out in the rapidity range $2.0 < y_{lab} < 3.0$ for a system of radius $R_o = 5.0$ fm, with $Z = 32$, $T = 200$ MeV, $\mu = 0$, and $p_z = 0$. The transverse momentum distribution of the charged pions is found to be quite sensitive to the rapidity of the pion. The pion transverse momentum distribution from $p + p$ collisions is parametrized by [28]

$$\frac{1}{N_{event}} \frac{1}{p_T} \frac{dN}{dp_T} \approx e^{-\frac{1}{B}\sqrt{p_T^2 + C^2}}, \tag{19}$$

where $B = 0.162 \pm 0.003$ GeV and $C = 0.11 \pm 0.02$ GeV, and the normalization is defined by

$$\int_{p_T=0}^{2\,GeV} \frac{dN}{d\vec{p}_T} d\vec{p}_T = 1.$$

Fig. 7 shows the ratios of the normalized pion transverse momentum distribution relative to that from $p + p$ collisions at comparable energies for a pion system with a boundary (solid curve) and for the Bose-Einstein distribution (dashed curve), in comparison with experimental data for O+Au $\rightarrow negatives + $ X [28]. The Coulomb final-state interaction given by Eq. (18) is again included, and calculations were carried out for the rapidity range $2.0 < y_{lab} < 3.0$ for a system of radius $R_o = 7.0$ fm, with $Z = 60$, $T = 200$ MeV, $\mu = 0$, and $p_z = 0$.

The last two figures, Fig. 8 and Fig. 9, show a direct comparison of these pion transverse momentum distributions, for the case with a boundary (Eq. (12), Figs. 9 (a) and 8 (a)) and Bose-Einstein (Eq. (13), Figs. 9 (b) and 8 (b)) with data from S+S $\rightarrow negatives + $ X and O+Au $\rightarrow negatives + $ X [6–8], taking into account the Coulomb final-state interaction. Both sets of data were fitted using $T = 200$ MeV, $\mu = 0$, $2.0 < y_{lab} < 3.0$, and $R_o = 5.0$ fm and $Z = 32$ for S+S data and $R_o = 7.0$ fm and $Z = 60$ for O+Au data. It is clear that incorporation of both the boundary effects and the Coulomb final-state interaction effects on the pion system leads to good agreement with the data over the range of $p_T$ under consideration, as shown in Figs. 9 (a) and 8 (a). Also, Bose-Einstein distribution along with Coulomb final-state interactions give good agreement with data and it give a slightly larger enhancement at low-$p_T$, as in Figs. 9 (b) and 8 (b).

Coulomb final-state interactions predict also that there are significant differences between the transverse momentum distributions of $\pi^+$, $\pi^-$, and $\pi^o$ in the low transverse momentum region.

## VI. SUMMARY AND CONCLUSIONS

The effect of a boundary on a system of massless bosons (gluons) and a system of massive bosons (pions) at high temperature have been studied. The boundary is found to affect the transverse momentum distribution of these systems. Relative to a free-particle Bose-Einstein distribution, the transverse momentum distribution of gluons and pions is suppressed at low momenta and enhanced at high momenta. For the gluons in a quark-gluon plasma of a radius of only a few times $T^{-1}$, the transverse momentum distribution at low momenta relative to $R_o^{-1}$ is much smaller than the Bose-Einstein distribution and approaches it as the radius of the plasma goes to infinity. This suppression is due to the lower level density of single-particle states at the bottom of the well for a finite system [29]. It can also be depicted as due to the weakness of large-wavelength amplitudes in a bound system. In contrast, the high momentum tail arises from the rapid variation of the single-particle wave function at the boundary of the system. It is clear that the boundary effects on a quark-gluon plasma play an important role in the gluon transverse momentum distribution, and consequently on signatures of the plasma. The boundary effects should be taken into account to provide more reliable estimates of the signatures of quark-gluon plasma-formation.

The pions in a hadronic system produced after a high energy nucleus-nucleus collision interact with each other and with the surrounding hadronic and nuclear matter. These interactions, in the isospin state $I = 0$, are attractive. The pions can be modeled as moving in a mean-field potential which creates an effective boundary for the system. Here we assume that the pion system has a sharp boundary before freeze-out. The general features of the pion transverse momentum distribution are rather similar to that of the gluons. There is a suppression at low momenta, with a scale that depends on the radius of the system, and there is a large enhancement relative to Bose-Einstein distribution at high momenta, $p_T > 2.0$ GeV. The distribution is quite different from Bose-Einstein for systems of small radius, and approaches Bose-Einstein as the radius is taken to infinity. However, the pion transverse momentum distribution is broader than that of gluons at low momenta. For $p_T \geq m_\pi$, good agreement with experimental data is obtained for the pion transverse momentum distributions for S+S $\rightarrow negatives + $ X and O+Au $\rightarrow negatives + $ X .

The pion transverse momentum distribution at low $p_T$ for a system with a boundary does not depart considerably from a Bose-Einstein distribution for large systems, when compared with the available experimental data. However, the bounded model predicts that there is always an enhancement relative to the Bose-Einstein distribution at high momenta, $p_T \geq 2.0$ GeV. There is a possible contamination from hard scattering processes to the high transverse momentum tail, which makes the boundary effect difficult to observe. However, structure function is poorly known in this $p_T$ region and a quantitative study is needed in the future to differentiate between the two mechanisms.

What is the momentum distribution of an expanding pion system. The transverse momentum distribution of the pions is suppressed at low $p_T$ (relative to static system) giving rise to a depression in the distribution in that region, if the particles in the system have been boosted by the same amount. However, if the boost increases with the energy of the single particle wave function, a very small suppression in the transverse momentum distributions at low $p_T$ is found and the magnitude of the suppression is slightly increases with the amount of the boost. On the other hand, we have found no change in the high transverse momentum tail with the magnitude of the boost, while the intermediate portion of the distribution slightly increases. A pion system with a boost behaves as if it is approximately a system with a higher temperature.

The puzzle of the pion transverse momentum distribution in the low $p_T$ region, $p_T \leq m_\pi$, where an enhancement is observed in the data, leads us to study the effect of Coulomb final-state interactions on the system. Coulomb final-state interaction on pions arise from target and projectile remnants near the central rapidity region. This Coulomb final-state interaction modifies the charged pion transverse momentum distributions considerably at low transverse momenta, where the distribution of charged pions relative to neutral pions is enhanced for $\pi^-$ but suppressed for $\pi^+$. The degree of enhancement or suppression is quite sensitive to the rapidity of the particles in the system. The inclusion of effects calculated from Coulomb final-state interactions leads to very good agreement with $\pi^-$ experiment. differences between the transverse momentum distributions of $\pi^+$, $\pi^-$, and $\pi^o$ in the low transverse momentum region await future experimental data for direct comparison.

### ACKNOWLEDGMENTS

The authors would like to thank Dr. T. Barnes and Dr. Y. Takahashi for helpful discussions. One of the authors (M.M.) would like to thank Prof. A. Goneid for helpful discussions, Ain Shams University for financial support, and Dr. M. Strayer and Dr. J. Ball for their kind hospitality at ORNL. This research was supported in part by the Division of Nuclear Physics, U.S. Department of Energy under Contract No. DE-AC05-84OR21400 managed by Martin Marietta Energy Systems, Inc.

# REFERENCES

[1] For an introduction and a review of current research in the area of quark-gluon plasma see: C.-Y. Wong, *Introduction to High-Energy Heavy-Ion Collisions* (World Scientific, 1994); Proceedings of Quark Matter '91, Gatlinburg, U.S.A., 1991, published in Nucl. Phys. **A544** (1991); Proceedings of the Quark Matter '93, Borlänge, Sweden, published in Nucl. Phys. **A566** (1994).
[2] R. C. Hwa and K. Kajantie, Phys. Rev. **D32**, 1109 (1985).
[3] K. Kajantie, J. Kapusta, L. McLerran, and A. Mekjian, Phys. Rev. **D34**, 2746 (1986).
[4] P. V. Ruuskanen, Nucl. Phys. **A522**, 255c (1991).
[5] C.-Y. Wong, Phys. Rev. **C48**, 902 (1993).
[6] H. Ströbele (NA35 Collab.), Z. Phys. **C38**, 89 (1988).
[7] J. W. Harris (NA35 Collab.), Nucl. Phys. **A498**, 133 (1989).
[8] R. Renfordt (NA35 Collab.), Nucl. Phys. **A498**, 385 (1989).
[9] T. W. Atwater, P. S. Freier, and J. Kapusta, Phys. Lett. **B199**, 30 (1987).
[10] K. S. Lee and U. Heinz, Z. Phys. **C43**, 425 (1989); K. S. Lee, U. Heinz, and E. Schnedermann, Z. Phys. **C48**, 525 (1990).
[11] D. Kusnezov and G. Bertsch, Physics Rev. **C40**, 2075 (1989).
[12] M. Kataja and P. V. Ruuskanen, Phys. Lett. **B243**, 181 (1990).
[13] P. Gerber, H. Leutwyler, and J. L. Goity, Phys. Lett. **B246**, 513 (1990).
[14] S. Gavin and P.V. Ruuskanen, Phys. Lett. **B262**, 326 (1991).
[15] S. Sollfrank, P. Koch, and U. Heinz, Phys. Lett. **B252**, 256 (1990).
[16] H. W. Barz, G. Bertsch, D. Kusnezov, and U. Heinz, Phys. Lett. **B254**, 332 (1991).
[17] J. Bolz, U. Ornik, and R. M. Weiner, Phys. Rev. **C46**, 2047 (1992).
[18] *Handbook of Mathematical Functions*, edited by M. Abramowitz and I. A. Stegun (Dover Publications, New York, 1972).
[19] Yu.M. Sinyukov, Nucl. Phys. **A566** 589c (1994).
[20] M. Kataja, P.V. Ruuskanen, L.D. McLerran, and H. von Gersdorff, Phy. Rev. **D34**, 2755 (1986).
[21] T. Barnes and E. S. Swanson, Phys. Rev **D46**, 131 (1992).
[22] J. D. Bjorken, Phys. Rev.**D27**, 140 (1983).
[23] E. V. Shuryak, Phys. Rev. **D42**, 1764 (1990).
[24] P. Seyboth (NA35 Collab.), Nucl. Phys. **A544**, 293 (1992).
[25] W. Beneson *et al.*, Phys. Rev. Lett. **43**, 683 (1979); Erratum: Phys. Rev. Lett. **44**, 54 (1980).
[26] K. G. Libbrecht and S. E. Koonin, Phys. Rev. Lett. **43**, 1581 (1979).
[27] M. Gyulassy and S. K. Kaufmann, Nucl. Phys. **A362**, 503 (1981).
[28] D. Röhrich (NA35 Collab.), Nucl. Phys. **A566**, 35c (1994).
[29] M.Brock, J.Damgaard, A.S.Jensen, H.C. Pauli, V.M. Strutinsky, and C.-Y. Wong, Rev. Mod. Phys. **44**, 320 (1972).

FIGURES

FIG. 1. The transverse momentum distributions of gluons in a quark-gluon plasma with a sharp cylindrical boundary, for radii $R_o = 2$, 3.5, and 5 fm, at temperature $T = 200$ MeV, $\mu = 0$, $m_g = 0$, and $p_z = 0$. The solid curves are the numerical results obtained using Eq. (10), and the dashed curves are the predictions for a Bose-Einstein distribution, given by Eq. (11).

FIG. 2. The transverse momentum distributions of pions in a pion system with a sharp cylindrical boundary, for radii $R_o = 2$, 4, and 6 fm, at temperature $T = 200$ MeV, $\mu = 0$, $m_\pi = 140$ MeV, and $p_z = 0$. The solid curves are the distributions arising from the boundary effect, given by Eq. (12). The dashed curves are the predictions for a Bose-Einstein distribution, given by Eq. (13).

FIG. 3. The pion transverse momentum distributions in a pion system with a boundary given by Eq. (12) (solid curves), and the predictions from the Bose-Einstein distribution given by Eq. (13) (dashed curves), in comparison with the experimental data for S+S $\rightarrow$ $negatives$ + X at $E = 200A$ GeV [6–8]. Good agreement is obtained with a radius $R_o = 5$ fm, temperature $T = 200$ MeV, $\mu = 0$, $m_\pi = 140$ MeV, and $p_z = 0$.

FIG. 4. The pion transverse momentum distributions from Eq. (12) for a bounded pion system (solid curves), and a Bose-Einstein distribution (Eq. (13), dashed curves), in comparison with experiment for O+Au $\rightarrow$ $negatives$ + X at $E = 200A$ GeV [6–8]. Good agreement is obtained with a radius $R_o = 7$ fm, temperature $T = 200$ MeV, $\mu = 0$, $m_\pi = 140$ MeV, and $p_z = 0$.

FIG. 5. The effect of the Coulomb final-state interaction, calculated using Eq. (18) on the transverse momentum distributions of $\pi^+, \pi^-,$ and $\pi^o$ in a bounded pion system produced after the collision S+S $\rightarrow$ $negatives$ + X (see text).

FIG. 6. The ratios, relative to that of a $p+p$ collisions, of the normalized transverse momentum distributions of negatively charged pions in a pion system with a cylindrical boundary of radius $R_o = 5.0$ fm, given by Eq. (12) (solid curve), and the Bose-Einstein distribution, Eq. (13) (dashed curve), in comparison with data from the collision S+S $\rightarrow$ $negatives$ + X [28]. For the bounded pion system results have been calculated taking into account Coulomb final-state interaction, as in Eq. (18) (see text).

FIG. 7. The ratios of the normalized transverse momentum distributions of negatively charged pions in a pion system with a cylindrical boundary of radius $R_o = 7.0$ fm (Eq. (12), solid curve), and the Bose-Einstein distribution (Eq. (13), dashed curve), relative to that of $p + p$ collisions, in comparison with data from the collision O+Au $\rightarrow$ $negatives$ + X [28]. For the bounded pion system results have been calculated taking into account Coulomb final-state interaction, as in Eq. (18) (see text).

FIG. 8. The pion transverse momentum distributions in a pion system with a boundary, Eq. (12), as in Eq. (18) (fig.(a)), and the predictions from Bose-Einstein distribution, Eq. (13) (fig.(b)), taking into account Coulomb final-state interaction, in comparison with experiment for S+S $\rightarrow$ $negatives$ + X [6–8]. Theoretical results are obtained with radius $R_o = 5$ fm, for $Z = 32$, temperature $T = 200$ MeV, $\mu = 0$, $m_\pi = 140$ MeV, and $2.0 < y < 3.0$.

FIG. 9. The pion transverse momentum distributions in a pion system with a boundary, Eq. (12), as in Eq. (18) (fig.(a)), and the predictions from Bose-Einstein distribution, Eq. (13) ( fig.(b)), taking into account Coulomb final-state interaction, in comparison with experiment for O+Au $\rightarrow$ $negatives$ + X [6–8]. Theoretical results are obtained with radius $R_o = 7$ fm, for $Z = 60$, temperature $T = 200$ MeV, $\mu = 0$, $m_\pi = 140$ MeV, and $2.0 < y < 3.0$.